\documentclass{pasj00}
\begin{document}	
\SetRunningHead{M. Ozawa et al.}{Energy-Scale Calibration of the XIS}
\Received{2008/08/15}
\Accepted{2008/09/20}
\title{Energy-Scale Calibration of the Suzaku X-Ray Imaging Spectrometer \\ Using the Checker Flag Charge Injection Technique in Orbit}
\author{
Midori \textsc{Ozawa},\altaffilmark{1}
Hideki \textsc{Uchiyama},\altaffilmark{1}
Hironori \textsc{Matsumoto},\altaffilmark{1}
Hiroshi \textsc{Nakajima},\altaffilmark{2}
Katsuji \textsc{Koyama},\altaffilmark{1} \\
Takeshi Go \textsc{Tsuru},\altaffilmark{1} 
Masahiro \textsc{Uchino},\altaffilmark{2}
Hiroyuki \textsc{Uchida},\altaffilmark{2}
Kiyoshi \textsc{Hayashida},\altaffilmark{2}
Hiroshi \textsc{Tsunemi},\altaffilmark{2}  \\
Hideyuki \textsc{Mori},\altaffilmark{3} 
Aya \textsc{Bamba},\altaffilmark{3}
Masanobu \textsc{Ozaki},\altaffilmark{3}
Tadayasu \textsc{Dotani},\altaffilmark{3} \\
Takayoshi \textsc{Kohmura}\altaffilmark{4} 
Yoshitaka \textsc{Ishisaki}\altaffilmark{5}
Hiroshi \textsc{Murakami},\altaffilmark{6} \\
Takeshi \textsc{Kato},\altaffilmark{7} 
Takeshi \textsc{Kitazono},\altaffilmark{7}
Yuki \textsc{Kimura},\altaffilmark{7}
Kazuki \textsc{Ogawa},\altaffilmark{7}
Shusuke \textsc{Kawai},\altaffilmark{7}
Koji \textsc{Mori},\altaffilmark{7} \\
Gregory \textsc{Prigozhin},\altaffilmark{8} 
Steve \textsc{Kissel},\altaffilmark{8}
Eric \textsc{Miller},\altaffilmark{8}
Beverly \textsc{LaMarr},\altaffilmark{8}
and
Marshall \textsc{Bautz},\altaffilmark{8}} 
\altaffiltext{1}{Department of Physics, Kyoto University, Kita-Shirakawa, Sakyo-ku, Kyoto, 606-8502}
\email{midori@cr.scphys.kyoto-u.ac.jp}
\altaffiltext{2}{Department of Earth and Space Science, Osaka University, 1-1 Machikaneyama, Toyonaka, Osaka 560-0043}
\altaffiltext{3}{Institute of Space and Astronautical Science, JAXA, Yoshinodai, Sagamihara, Kanagawa 229-8510}
\altaffiltext{4}{Department of General Education, Kogakuin University, 2665-1 Nakano-Cho, Hachioji, Tokyo 192-0015}
\altaffiltext{5}{Department of Physics, Tokyo Metropolitan University, 1-1 Minami-Osawa, Hachioji, Tokyo 192-0397}
\altaffiltext{6}{Department of Physics, Rikkyo University, 3-34-1 Nishi-Ikebukuro, Toshima-ku, Tokyo 171-8501}
\altaffiltext{7}{Department of Applied Physics, University of Miyazaki, 1-1 Gakuen Kibana-dai Nishi, Miyazaki 889-2192}
\altaffiltext{8}{Center for Space Research, Massachusetts Institute of Technology, Cambridge, MA 02139-4307, USA}
\KeyWords{instrumentation: detectors ---techniques: spectroscopic ---X-ray CCDs} 
\maketitle

\begin{abstract}
The X-ray Imaging Spectrometer (XIS) on board the Suzaku satellite is an X-ray CCD
camera system that has superior performance such as a low background, high 
quantum efficiency, and good energy resolution in the 0.2--12~keV band. 
Because of the radiation damage in orbit, however, the charge transfer inefficiency 
(CTI) has increased, and hence the energy scale and resolution of the XIS has
been degraded since the launch of July 2005.  The CCD has a charge injection 
structure, and the CTI of each column and the pulse-height dependence of the CTI are precisely 
measured by a checker flag charge injection (CFCI) technique. Our precise CTI 
correction improved the energy resolution from 230~eV to 190~eV at 5.9~keV in December 2006.  This paper reports the CTI measurements with the CFCI 
experiments in orbit.  Using the CFCI results, we have implemented 
the time-dependent energy scale and resolution to the Suzaku calibration database.
\end{abstract}

\section{Introduction} 
After the first successful space flight use of the X-ray charge coupled device (CCD) of 
the SIS (\cite{Burke93}) on board ASCA, the CCD has been playing a major role in imaging
spectroscopy in the field of X-ray astronomy.  However, the charge transfer inefficiency (CTI) of 
X-ray CCDs increases in orbit due to the radiation damage; the CTI is defined as
the fraction of electrons that are not successfully moved from one CCD pixel to the next
during the readout.  Since the amount of charge loss depends on the number of the
transfers, the energy scale of X-ray CCDs depends on the location of an X-ray event.
Furthermore, there is a fluctuation in the amount of the lost charge. Therefore, 
without any correction, the energy resolution of X-ray CCDs in orbit gradually degrades. 
In the case of the X-ray Imaging Spectrometer (XIS)~\citep{Koyama07} on board 
the Suzaku satellite~\citep{Mitsuda07} launched on July 10, 2005, the 
energy resolution in full width at half maximum (FWHM) at 5.9~keV was 
$\sim$140 eV in August 2005, but had degraded to $\sim$230 eV in December 2006.

The increase of the CTI is due to an increase in the number
of charge traps at defects in the lattice structure of
silicon made by the radiation. Since the trap distribution
is not uniform, it would be best if we could measure the CTI
of each pixel as Chandra ACIS~\citep{Grant04}. In the case
of the XIS, however, it is impossible to measure the CTI
values of all the pixels, mainly because the onboard
calibration sources do not cover the entire field of view of
the XIS. Therefore, we use the CTI of each column to correct
the positional dependence of the energy scale.

The XIS is equipped with a charge injection 
structure~\citep{Prigozhin04,Bautz04,LaMarr04,Prigozhin08}
which can inject an arbitrary amount of charge in arbitrary positions.  
Using this capability, we can precisely measure the CTI of each
column~\citep{Nakajima08}.  By applying the column-to-column
CTI correction, the positional dependence of the CTI corrected energy scale is greatly reduced,
and the over-all energy resolution is also improved~\citep{Nakajima08}.

In \citet{Nakajima08}, the results of the CTI correction was mainly based 
on the ground-based charge injection experiments.
In-orbit measurements were limited within one year after the launch. 
This paper reports more comprehensive and extended  in-orbit experiments 
up to two years after the launch.  
The results are based on the data with the normal full window mode \citep{Koyama07} 
without a spaced-row charge injection\footnote{After October 2006, the spaced-row charge 
injection \citep{Uchiyama09} has been a normal observation mode, and hence we should use different correction method.} \citep{Uchiyama09}, 
and have been implemented to the Suzaku calibration database and applied to all the data obtained with the same mode.  
All the errors are at the 1$\sigma$ confidence level throughout this paper unless mentioned.

\section{X-Ray CCD with the Charge Injection Capability} 
\subsection{Checker Flag Charge Injection \label{sec:CFCI}}
 
The XIS is the set of four X-ray CCD camera systems.  Three
sensors (XIS~0, 2, and 3) contain front-illuminated (FI)
CCDs and the other (XIS~1) contains back illuminated (BI)
CCD. The XIS~2 sensor became unusable on November 9,
2006. Therefore there are no data for XIS~2 after that day.

The detailed structure of the CCD has been provided in
\citet{Koyama07}.  In this paper, we define a "row" and a
"column" as a CCD line along the $ActX$ and $ActY$ axes,
respectively (see figure~3 in \cite{Koyama07}).  In the
imaging area, the \textit{ActX} value runs 0 to 1023 from
the segment A to D, while the \textit{ActY} value runs from
0 to 1023 from the readout node to the charge injection
structure.

The charge injection structure lies adjacent to the top row
(\textit{ActY} = 1023) in the imaging area. We can inject
charges from $\sim$ 50 e$^-$ to $\sim$ 4000 e$^-$ per pixel;
the equivalent X-ray energy ranges from $\sim$0.2~keV to
$\sim$15~keV.

A charge packet generated by an incident X-ray is
transferred to the readout node, then is converted to a
pulse-height value.  We define $PH_{\rm o}$ to be the
original pulse height generated by the X-ray. In the real
case, the readout pulse height of the packet ($PH'$) is
smaller than $PH_{\rm o}$, because some amount of charges is
lost during the transfer.  To measure the charge loss, we
have to know both $PH_{\rm o}$ and $PH' $.  However, we can
usually measure only $PH'$, and hence it is difficult to
obtain $PH_{\rm o}$.

\citet{Prigozhin04} and \citet{Prigozhin08} reported a technique to solve this problem by the charge injection method,
and \citet{Nakajima08} applied this technique to the XIS.
We briefly repeat by referring figure~3 in
\citet{Nakajima08}.  First, we inject a ``test'' charge
packet into the top CCD row (\textit{ActY} = 1023).  Then,
after the gap of a few rows, five continuous packets are
injected with the same amount of charge of the test
packet. The former four packets are called ``sacrificial''
charge packets, while the last one is called a ``reference''
charge packet.  The test packet loses its charge by the
charge traps. On the other hand, the reference packet does
not suffer from the charge loss, because the traps are
already filled by the preceding sacrificial packets. Thus we
can measure the charge loss by comparing the pulse-height
values of the reference charge ($PH_{\rm ref}$) and the test
charge ($PH_{\rm test}$). The relation between sacrificial charge packets and 
reference charge packets is described in Gendreau (1995). 
We can obtain a checker flag
pattern by these injected packets in the X-ray image (right
panel of figure~3 in \cite{Nakajima08}) because of the
onboard event-detection algorithm~\citep{Koyama07}.
Therefore in this paper, we call this technique a ``checker
flag charge injection (CFCI).''

\subsection{Formulation of the CTI\label{sec:formulation}}

A charge packet in the XIS loses its charge during (a) the
fast transfer (24~$\mu$s~pixel$^{-1}$) along the {\it ActY}
axis in the imaging area, (b) the fast transfer along the
{\it ActY} axis in the frame-store region, (c) the slow
transfer (6.7~ms~pixel$^{-1}$) along the {\it ActY} axis in
the frame-store region, (d) the fast transfer to the readout
node along the {\it ActX} axis. The CTI depends on many
parameters such as the transfer speed and the number density
of the charge traps~\citep{Hardy98}.  The frame-store region
is covered by the shield and is not exposed to the radiation
directly.  Furthermore, the pixel size of the frame-store
region (21~$\mu$m$\times$13.5~$\mu$m) is different from that
of the imaging area (24~$\mu$m$\times$24~$\mu$m). Thus the
number of traps per pixel may be different between the
imaging area and the frame-store region.
Then we assumed that the four transfers have different CTI values. 
We examined the transfer (d) by using the calibration source data taken 
in April 2007, and found
no significant decrease of the pulse height along the {\it ActX} axis. 
We, therefore, ignore the 
charge loss in the transfer (d). 

We define that $i$ is the transfer number in the imaging area 
($i$=$ActY$+1; here, $ActY$ is a coordinate value where 
an incident X-ray generates a charge packet).
Then the relation between $PH'$ and $PH_{\rm o}$  is expressed as, 
\begin{eqnarray}
PH'(i) &=& PH_{\rm o} (1 - c_a)^i (1 - c_b)^{1024-i} (1 - c_c)^i, \nonumber \\
&\sim& PH_{\rm o} \left\{1 - i (c_a - c_b + c_c)  - 1024 c_b \right\}, 
\label{eq:cti_def_0}
\end{eqnarray}
where $c_a$, $c_b$ and $c_c$ are the CTI values in the transfers (a), (b), and (c), respectively.  
Here we used the fact that the CTI values are much smaller than 1.  
Thus we can separate the charge loss into $i$-dependent component (the second term in the 
right-hand side of equation~\ref{eq:cti_def_0}) and constant component (the third term).
We therefore substitute the CTI with CTI\,1 (the former component) and CTI\,2 (the latter component), 
which have the CTI values of $c_1 = c_a - c_b + c_c$ and $c_2 = c_b$, respectively.
Then equation~\ref{eq:cti_def_0} can be written as
\begin{equation}
PH'(i) \sim PH_{\rm o} (1 - i c_1 - 1024 c_2).
\label{eq:cti_def}
\end{equation}

Since the CTI values depend on the amount of transfer
charge which is proportional to the pulse height, 
we assume the CTI is described by a power function of the pulse height (Prigozhin et al. 2004)
and expressed as
\begin{equation}
c_{1} = k_{1} (PH_{\rm o})^{-\beta} \hspace{1em} {\rm and} \hspace{1em}
c_{2} = k_{2} (PH_{\rm o})^{-\beta},
\label{eq:cti_ph}
\end{equation}
where $k_1$ and $k_2$ are scale factors for the
CTI\,1 and CTI\,2, and the index $\beta$ is common to the CTI\,1 and CTI\,2. 

\section{Measuring the CTI with the CFCI\label{sec:measure}} 

We have conducted the CFCI experiments six times in orbit.  
Effective exposure time for each experiment ranges from a few to $\sim$20 ks.  
The equivalent X-ray energy of the injected charge packets
ranges from $\sim$0.3~keV to $\sim$8~keV.  Since June 2006,
we injected various amounts of charge in one experiment.
The log is summarized in table~1.

In the CFCI experiments, the test charge is injected to the
row at $\textit{ActY} = 1023$ ($i = 1024$), and hence
$PH'(1024) = PH_{\rm test}$.  The reference charge should be
equal to the original charge which does not suffer from the
charge loss, and hence $PH_{\rm o} = PH_{\rm ref}$. Then
equation~\ref{eq:cti_def} can be written as
\begin{equation}
c_1 + c_2 = \frac{1}{1024} \left( 1 - \frac{PH_{\rm test}}{PH_{\rm ref}}\right).
\label{eq:cfci_cti}
\end{equation}

\begin{figure}
\begin{center}
\FigureFile(80mm,80mm){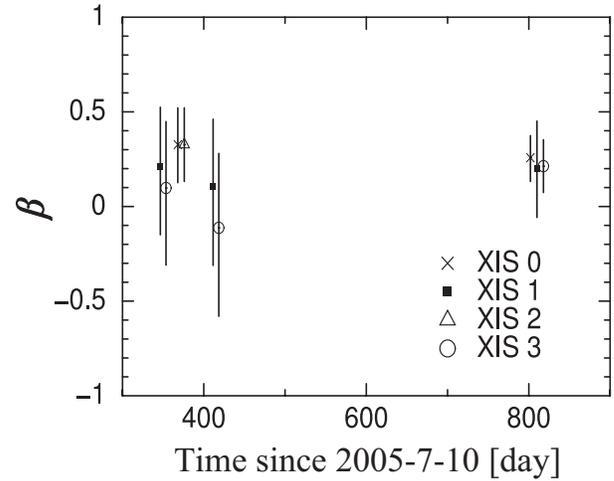}
\end{center}
\caption{Time evolution of $\beta$ averaged over one sensor.
We show the standard deviation of $\beta$ as the error bar.	
 \label{fig:ctipow}}
\end{figure}

We determined $c_1 + c_2$ by measuring the ratio $PH_{\rm test}/PH_{\rm ref}$ for each column.
From equation~3, we can obtain the relation in the CFCI experiments as,
\begin{equation}
c_1 + c_2 = (k_1+k_2)PH_{\rm ref}^{-\beta}.
\label{eq:cfci_cti2}
\end{equation}
The index $\beta$ and $k_1+k_2$ were derived by fitting
equation~\ref{eq:cfci_cti2} to the values of $c_1 + c_2$
obtained with the CFCI experiments with multiple amount of
charge injections (multiple $PH_{\rm ref}$s; Log Number 3--6
in table~1).  The mean and standard deviation of the
best-fit $\beta$ of equation~\ref{eq:cfci_cti2} averaged
over each sensor are shown in figure~\ref{fig:ctipow}. The
mean value of $\beta$ shows no time variation, and the time
averaged values of XIS~0, 1, and 3 are 0.31, 0.22, 0.15 for
XIS~0, 1, and 3, respectively.  As for XIS~2, there was only
one data point, and we obtained $\beta=0.34$.

If a charge packet has a volume proportional to the number
of electrons and is spherically symmetric, the probability
that one electron encounters a charge trap is proportional
to the cross section of the charge packet.  In this case, we
can expect $(PH_{\rm ref} -PH_{\rm test})\propto PH_{\rm
ref}^{2/3}$. From equation~\ref{eq:cfci_cti}, the CTI value
is proportional to $(PH_{\rm ref} -PH_{\rm test})/ PH_{\rm
ref}$, and hence $\propto PH_{\rm ref} ^{-1/3}$.  Thus the
simple model is roughly consistent with the observed $\beta$
values.

Using the results of all the CFCI experiments (Log Number 1--6 in table~1) and 
above determined  $\beta$ values, we then re-estimated $k_1$ and $k_2$ separately.
In this process, we assumed $k_1/k_2$ is equal to $c_1/c_2$ which was estimated 
by the 6.4~keV line from the Sgr~C region to be 0.67 and 1.5 for the FI and BI CCD, respectively (\cite{Nakajima08}).  From this $k_1$, we can 
obtain the final value of $c_1$.

\begin{figure}
\begin{center}
\FigureFile(80mm,80mm){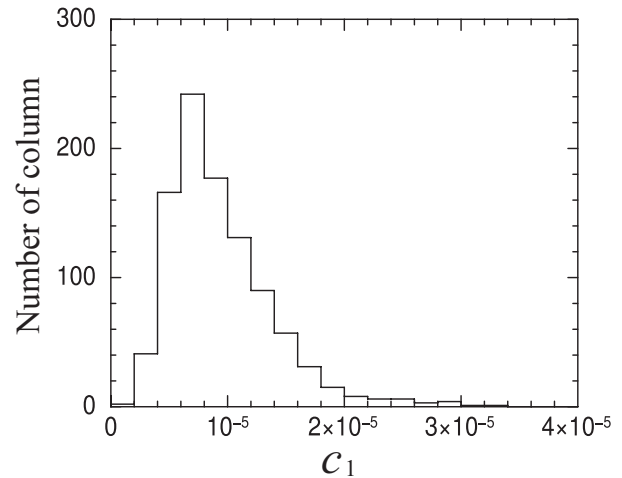}
\end{center}
\caption{The number distribution of the CTI\,1 parameter ($c_1$).
The parameter $c_1$ is obtained by the checker flag charge injection (CFCI) experiment on July 17, 2006.	
We show the result of XIS~0 as a typical example.
The typical error	 of $c_1$ is	$\sim$	4$\times 10^{-7}$.
The mean and the peak (the most probable) values are 9.4$\times$10$^{-6}$ and
6.8$\times$10$^{-6}$, respectively.
The equivalent X-ray energy of the reference charge packet is 4.2~keV.\label{fig:baratuki}}
\end{figure}

\begin{figure}
\begin{center}
\FigureFile(80mm,80mm){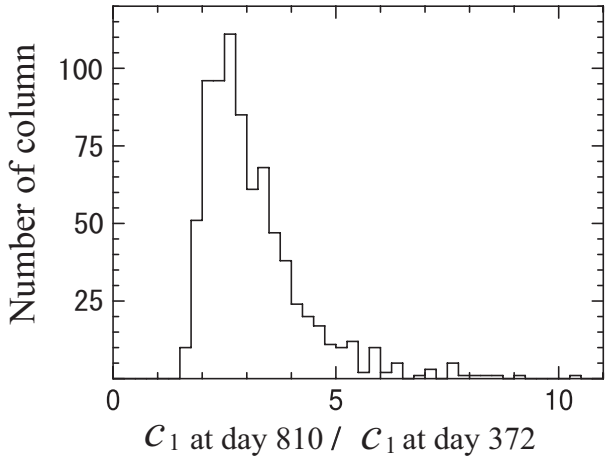}
\end{center}
\caption{Same as figure~2, but the ratio of $c_1$ obtained on July	17, 2006 (372
days after the launch) and on September 28, 2007 (810 days after the launch).	
The typical error of the ratio is $\sim$ 0.1.
The mean and the peak values  are 2.89 and 2.63, respectively.
Time evolution of $c_1$	(increase with time) is clearly seen.	
\label{fig:ctihenka}}
\end{figure}

\begin{figure}
\begin{center}
\FigureFile(80mm,80mm){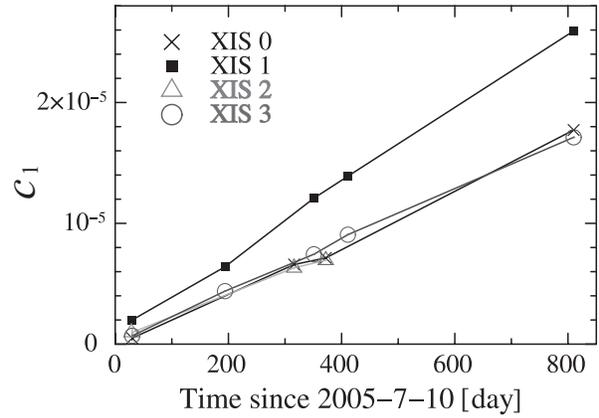}
\end{center}
\caption{Time history of column-averaged $c_1$ at 7.0~keV.
We obtained $c_1$ at day 32 (August 11, 2005) from the calibration
sources	and that at the other day from the CFCI experiments.
We used	the relation $c_{1} \propto (PH_{\rm o})^{-\beta}$	
when obtaining	$c_{1}$	at 7.0~keV. Here	we used	$\beta$	 shown	in section~3. \label{fig:ctijikanhatten}}
\end{figure}

Figure~\ref{fig:baratuki} shows an example of the distribution of $c_1$ in July 2006.  We can see significant 
column-to-column dispersion.
Figure~\ref{fig:ctihenka} shows the change of $c_1$ from July 2006 to September 2007.
We can see that the CTI values of all columns increased, but the increasing rate
was different from column to column. The results of figures~2 and 3 indicate that
the CTI correction at the column level is strongly required.
In figure~\ref{fig:ctijikanhatten}, we show the column-averaged $c_1$ value as a function of time.  

Since the CFCI experiments were only sparsely conducted (see table~1), 
we interpolate the $c_1$ and $c_2$ values for the observations of inter-CFCI epochs. 
As for the determination of the CTI values before the first CFCI experiment, see
appendix.

\section{Results of the CTI Correction and the Energy Calibration} 

A CTI correction, which is the conversion of $PH'(i)$ to
$PH_{\rm o}$, is made with equation~\ref{eq:cti_def}, where
$c_1$ and $c_2$ are calculated from equation~3 by using the
$k_1$, $k_2$, and $\beta$ values determined in section~3.

\subsection{Emission Lines for the Calibration}

We used the emission lines from the onboard calibration
sources, the Perseus cluster of galaxies, and the supernova
remnant 1E\,0102.2$-$7219.  We retrieved the data from the
Data Archives and Transmission System\footnote{
http://darts.isas.jaxa.jp/} of ISAS/JAXA.  All data were
acquired with the normal full window mode and the 3$\times$3
or 5$\times$5 editing mode~\citep{Koyama07}.  We used the
data with the ASCA grades of 0, 2, 3, 4, and 6.  As is
mentioned in \citet{Koyama07}, a small fraction of the
charge in a pixel is left behind (trailed) to the next pixel
in the same column during the transfer.  All data used in
this paper were corrected for the trail phenomenon. The
observations are summarized in table~\ref{tab:objlog}.

\subsection*{Onboard Calibration Source $^{55}$Fe}
The calibration source $^{55}$Fe produces the
Mn\emissiontype{I} K$\alpha$ line.  The theoretical line
center energy is 5895~eV \citep{Bearden67,Krause79}. 
We used the data from August 2005 to April 2007.  

\subsection*{The Perseus Cluster of Galaxies}
This is one of the X-ray brightest clusters of galaxies in the sky. 
The X-ray spectrum is that of a  thin thermal plasma with the strong K$\alpha$ line 
of Fe\emissiontype{XXV}.  The plasma temperature changes smoothly from $kT
\sim$ 4~keV to $\sim$ 7~keV toward the outer region~\citep{Chu03},
and the center energy of the Fe\emissiontype{XXV} K$\alpha$
triplet is almost constant ($\sim$6.56~keV at $z=0.0176$)
within this temperature range.  Its radius of $\sim 15'$ can
cover the entire field of view of the XIS ($18'\times 18'$). Thus this source is suitable 
for measuring the positional dependence of the energy scale.  

\subsection*{1E$\,0102.2-7219$}
This is one of the brightest supernova remnants in the
Small Magellanic Cloud. With the spatial resolution of
Suzaku, it can be regarded  as a point source.  
There are many bright emission lines originated
from thermal plasma in the X-ray spectrum below 2~keV.
These lines are resolved with the XMM-Newton RGS, 
and the accurate energies of the line centroids are
known~\citep{Ras01}.  This object has been used by many
instruments for the calibration in the low-energy band, and an
empirical model to describe the spectrum has been
established\footnote{http://cxc.harvard.edu/acis/E0102/}. 
We used this source as the energy-scale calibrator in the low-energy band.  

\subsection {Positional Dependence of the Energy Scale}

\begin{figure}
  \begin{center}
    \FigureFile(80mm,80mm){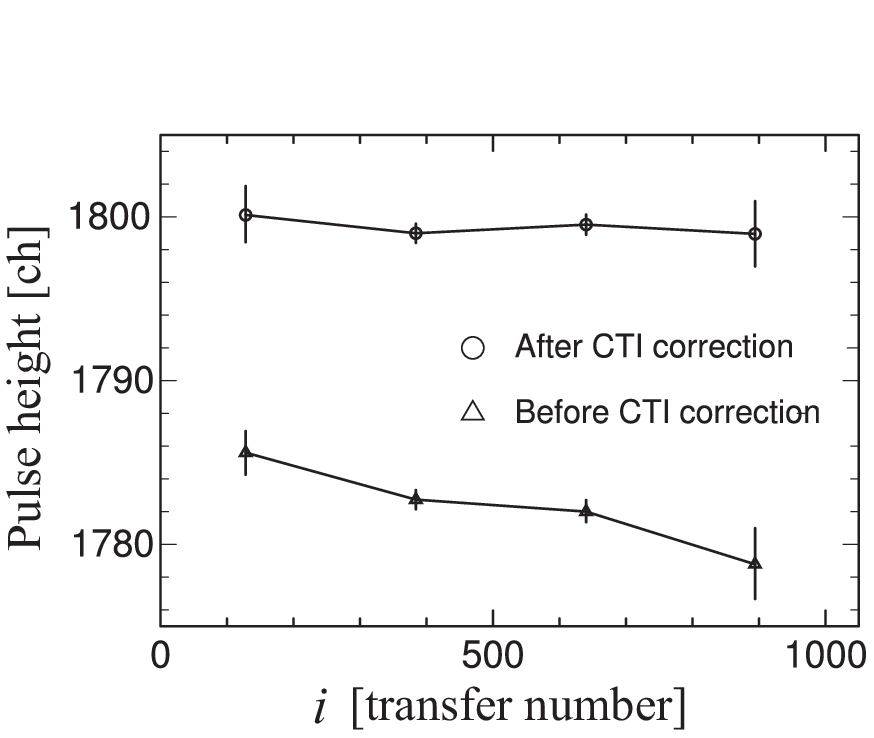}
  \end{center}
  \caption{Transfer number dependence of the center pulse height 
for the Fe\emissiontype{XXV} K$\alpha$ line of the Perseus cluster of galaxies. The result of XIS~0 is shown as a typical example.
Triangles and circles represent the data before and after 
the CTI correction, respectively. 
\label{fig:perseus}}
\end{figure}

For the data of the Perseus cluster of galaxies, 
we divided the imaging area into four regions along the {\it ActY} axis, and extracted a
spectrum from each region.  Then we fitted the spectra in
the 5--7.3~keV band with a power-law model and a
Gaussian function, and obtained the center pulse height of the
Fe\emissiontype{XXV} K$\alpha$ line.
Figure~\ref{fig:perseus} shows the center pulse height as a function
of $i$. Triangles and circles indicate the data before and
after the CTI correction, respectively. We can see no
significant $i$ dependence after the CTI correction, and
this supports the validity of our correction.

\subsection{Energy Scale}

The goal is to determine a relation of $PH_{\rm o}$ and
X-ray energy $E$. From the ground experiments, we found that
the $PH_{\rm o}$-$E$ relation can be expressed as a
broken-linear function linked at
the Si-K edge energy of 1839~eV~\citep{Koyama07}. 
We then determined the $PH_{\rm o}$-$E$ relation of each segment 
by using the lines of the calibration sources (Mn\emissiontype{I} K$\alpha$ line at 5895~eV) and 1E$\,0102.2-7219$  (K$\alpha$ lines of
O\emissiontype{VIII }, Ne\emissiontype{IX} and Ne\emissiontype{X} around 650--1020~eV).

\begin{figure}
  \begin{center}
    \FigureFile(80mm,80mm){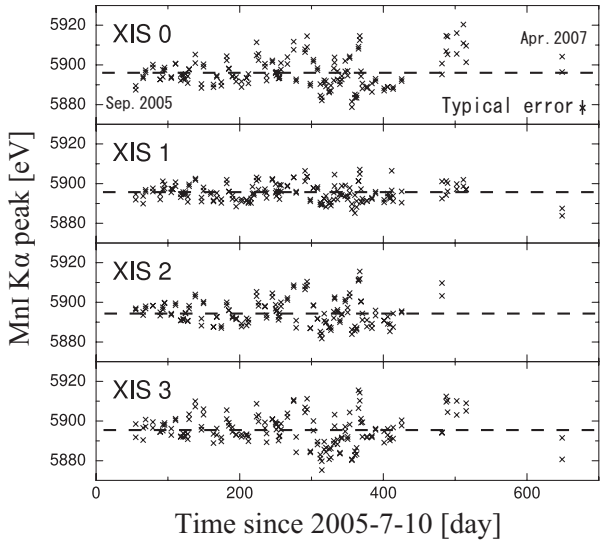}
  \end{center}
\caption{Center energy of the Mn\emissiontype{I} 
K$\alpha$ line for XIS~0--3 after the CTI correction. 
The theoretical center energy (5895~eV) is shown with dotted lines. 
Each mark in the plot has an effective exposure of more than 60~ks.
\label{fig:calpeak}}
\end{figure}

We show the results after the CTI correction and the $PH_{\rm o}$-$E$ conversion.
Figure~\ref{fig:calpeak} shows measured center energies of
the Mn\emissiontype{I} K$\alpha$ line as a function of time.
Each mark in the plot has an effective exposure of more than
60~ks.  The mean values of the center energy are 5896.2,
5895.4, 5895.0, and 5895.4 ~eV for XIS~0, 1, 2, and 3,
respectively.  The deviation around the theoretical center energy
(5895~eV) is 7.8, 4.4, 6.6, and 7.8~eV for XIS~0, 1, 2,
and 3, respectively.  Therefore, the time-averaged uncertainty of the
absolute energy is $\sim \pm$ 0.1 \% for the Mn\emissiontype{I} K$\alpha$ line of the calibration sources. 

\begin{figure}
  \begin{center}
    \FigureFile(80mm,80mm){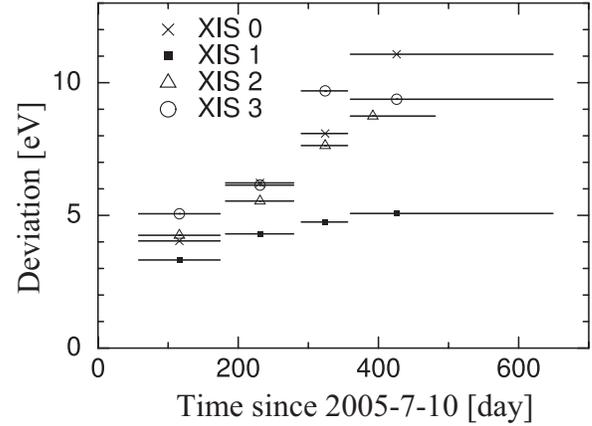}
  \end{center}
\caption{
The deviation of the center energy of the 
Mn\emissiontype{I} K$\alpha$ line  from the theoretical value (5895~eV)
as a function of time.
Each mark (except for the last one of XIS~2) is obtained from forty data points of figure~\ref{fig:calpeak}. 
\label{fig:peakbunsanhenka}}
\end{figure}

We also studied the time evolution of the deviation around the theoretical center energy, and
the results are shown in figure~\ref{fig:peakbunsanhenka}. We can see that the
deviation gradually increases with time.

\begin{figure}
\begin{center}
 \FigureFile(80mm,80mm){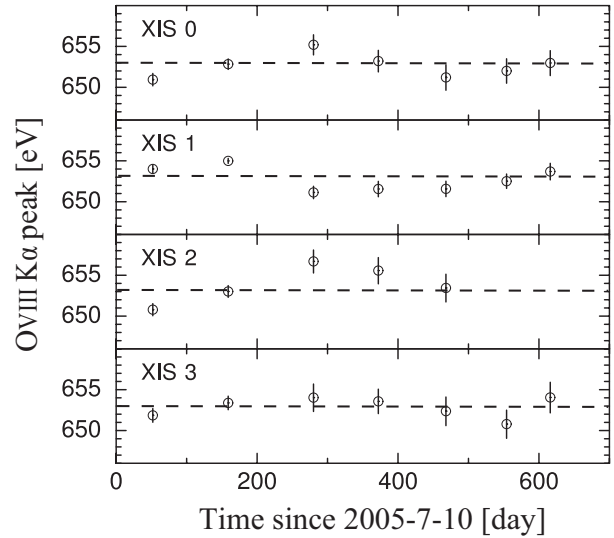}
\end{center}
\caption{
Center energy of the O\emissiontype{VIII}  K$\alpha$ line from 1E$\,0102.2-7219$ after the CTI correction.
Dotted lines indicate the center energy
of the empirical model (653~eV).\label{fig:sciofflowpeak}}
\end{figure}

Figure~\ref{fig:sciofflowpeak} shows the center energy of
the O\emissiontype{VIII} K$\alpha$ line from the 1E$\,0102.2-7219$ data.  
The mean values of the center energy are 652.6, 653.8, 652.7, and 652.8~eV for
XIS~0, 1, 2, and 3, respectively. The deviation around the center energy of the empirical model (653~eV) is
1.4, 1.4, 2.3, and 1.1~eV for XIS~0, 1, 2, and 3.
Therefore, the uncertainty of the absolute energy is $\sim
\pm$ 0.2\% for the O\emissiontype{VIII} K$\alpha$ line of 1E$\,0102.2-7219$.

\subsection{Energy Resolution}

We examined the energy resolution in FWHM ($\Delta E$) \citep{Koyama07} for each sensor; $\Delta E$ is common to all segments.
We expressed $\Delta E$ as
\begin{equation}
\Delta E~({\rm eV}) = \sqrt{ \left\{ a \times \left(\frac{E}{5895~{\rm eV}}\right) ^b\right\}^2 + (\Delta E_{\rm o
})^2},
\end{equation}
where $a$ and $b$ are time dependent parameters and 
$\Delta E_{\rm o}$ is the energy resolution determined by the ground experiments and obtained using equation~1 in \citet{Koyama07}. 
We determined $a$ and $b$ by using the
time history of the calibration sources and 1E$\,0102.2-7219$.  The $\Delta E$
values obtained in this way is incorporated into the
redistribution matrix file (RMF).

\begin{figure}
  \begin{center}
    \FigureFile(80mm,80mm){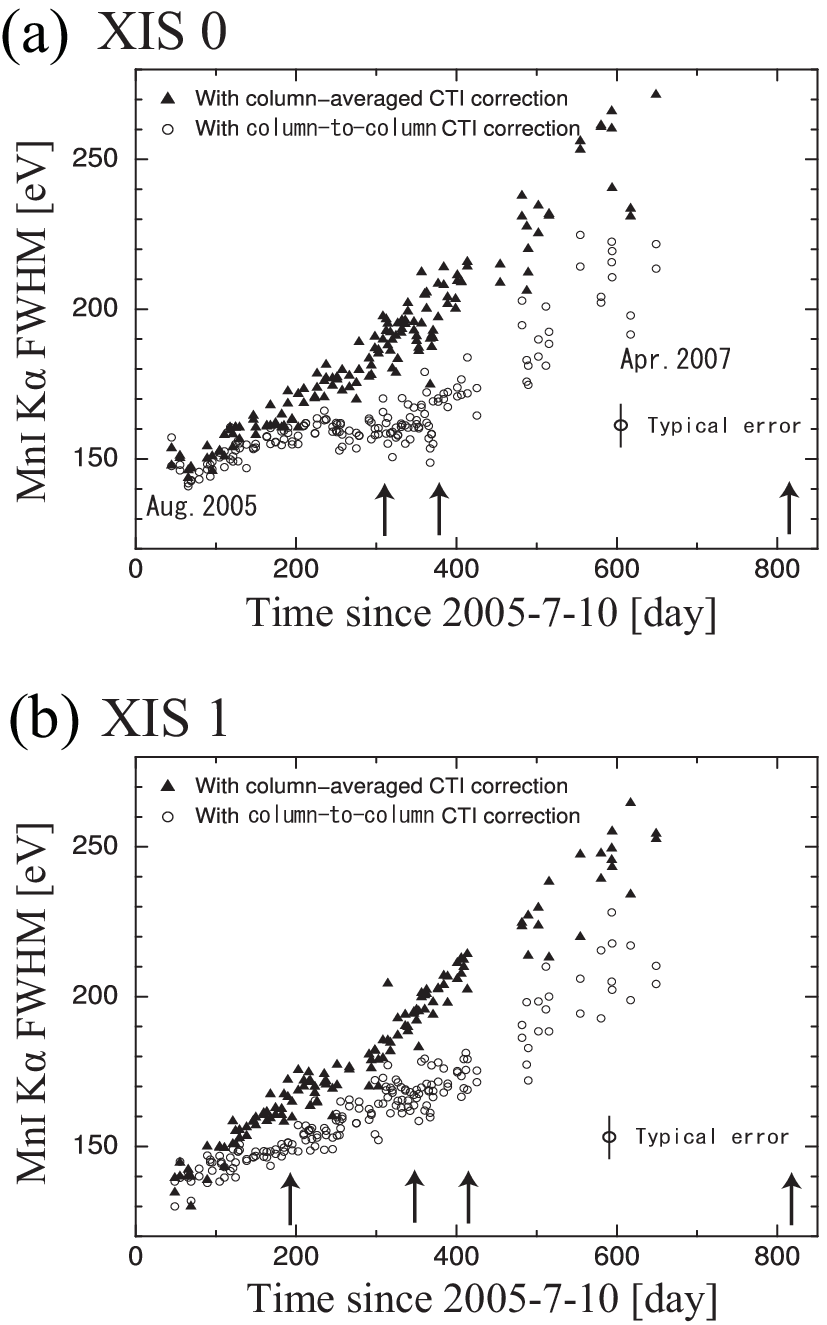}
  \end{center}
\caption{Comparison of the column-to-column CTI correction
(circles) and  the column-averaged CTI correction (triangles).
We show the energy resolution of the Mn\emissiontype{I} K$\alpha$ line of XIS~0 (a) and XIS~1 (b).
Arrows in the plot indicate the days we performed 
the CFCI experiments.\label{fig:segcolumn}}
\end{figure}

\begin{figure}
  \begin{center}
    \FigureFile(80mm,80mm){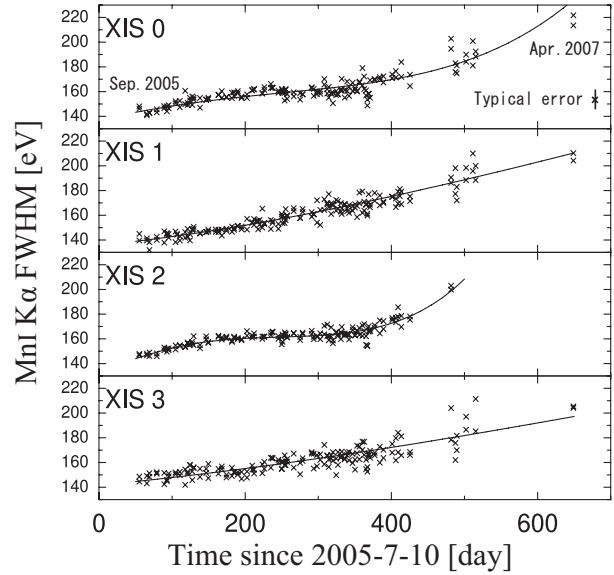}
  \end{center}
\caption{Energy resolution of the Mn\emissiontype{I} K$\alpha$
line for XIS~0--3 after the CTI correction. Each mark in the plot has an exposure time
of more than 60~ks. The solid lines indicate our model of
the energy resolution which is incorporated into the
redistribution matrix file. \label{fig:calsig}}
\end{figure}

Figure~\ref{fig:segcolumn} shows the energy resolution of the 
Mn\emissiontype{I} K$\alpha$ line after the
column-to-column CTI correction. We also plot the results
of the CTI correction, where we used the CTI values averaged
over a segment (the column-averaged CTI correction). 
We can see that the energy resolution is greatly improved by the
column-to-column CTI correction.
For example, the energy resolution in December 2006 was greatly
improved from $\sim$230~eV to $\sim$ 190~eV.
On the other hand, with the column-averaged CTI correction,
the energy resolution is $\sim$230~eV and is not significantly improved.  
In figure~\ref{fig:calsig},  we compared the energy resolution of
the Mn\emissiontype{I} K$\alpha$ line with our RMF model. The deviation
of the data points around our model is 5.6, 4.9, 3.4, and
6.3~eV for XIS~0, 1, 2, and 3, respectively.
\section{Summary} 

We have conducted the CFCI experiments six times in orbit. 
The CTI correction has been done with the CFCI results.
We calibrated the energy scale of the XIS precisely using the onboard calibration sources and 1E$\,0102.2-7219$.
Our calibration results have been applied to all the data  obtained with the normal full window mode without the spaced-row charge injection.
The results of the CFCI experiments and the current calibration status are 
summarized as follows:

\begin{enumerate}
\item 
We determined the CTI\,1 and CTI\,2 values of each column
precisely based on the data of the CFCI experiments.  We
also found that the pulse height dependence of the CTI does not change
with time.

\item
After the column-to-column CTI correction, we determined the
$PH_{\rm o}$-$E$ relation. We also modeled the time-dependent energy
resolution.

\item 
The uncertainty of the energy scale is $\pm$ 0.2 \% for the O\emissiontype{VIII} K$\alpha$ line ($\sim$ 0.65~keV) of 1E$\,0102.2-7219$,
and $\pm$ 0.1 \% for the Mn\emissiontype{I} K$\alpha$ line ($\sim $5.9~keV) of the calibration sources.

\item 
With the column-to-column CTI correction, the energy resolution at 5.9~keV in December 2006 was greatly improved from $\sim$230~eV to $\sim$ 190~eV.

\end{enumerate}

\bigskip
The authors thank all the XIS members for their support and useful information.
This work was supported by the Grant-in-Aid for the Global COE
Program "The Next Generation of Physics, Spun from
Universality and Emergence" from the Ministry of Education,
Culture, Sports, Science and Technology (MEXT) of Japan.
M.O., H.U., and H.N. are financially supported by the Japan
Society for the Promotion of Science.  H.M. is also supported
by the MEXT, Grant-in-Aid for Young Scientists~(B),
18740105, 2008, and by The Sumitomo Foundation, Grant for
Basic Science Research Projects, 071251, 2007.
H.T. and K.H. were supported by the MEXT, Grant-in-Aid 16002004.

\appendix

\section*{Determination of the CTI values before the first CFCI experiment
\label{app:cti0}}

First, we determined the CTI values of the segments A and D
in August 2005. We combined the data of
the calibration sources from August 11 to 31, 2005, and
obtained the $PH'$ of the Mn\emissiontype{I} K$\alpha$ line.
We also estimated $PH_{\rm o}$ at 5895~eV from the $PH_{\rm o}$-$E$ relation determined by the ground experiments, and
obtained the ratio $PH'/PH_{\rm o}$.  From
equation~\ref{eq:cti_def}, the ratio can be expressed
approximately as $PH'/PH_{\rm o} = 1 - i_{\rm cal} c_1 -
1024 c_2$, where $i_{\rm cal}$ is the mean transfer number of
the calibration events (typically $\sim$900). We determined $c_1$ and
$c_2$ at 5895~eV from $PH'/PH_{\rm o}$ with the $c_1/c_2$ ratio fixed
to the values shown in section~3.  The $c_1$
and $c_2$ for other pulse-height values were calculated from
equation~\ref{eq:cti_ph}, where we used the column-averaged and time-averaged $\beta$ values determined in section~3.  
Then for segments B and C, we took the average CTI values of the segments A and D.  
We regard the CTI values obtained in this procedure
as those on August 11, 2005 (the day of the XIS first light).  
Note that these values are determined for each segment, not for each column.

\begin{table*}
\caption{Log of the checker flag charge injection experiments in orbit.
\label{tab:cfcilog}}
\begin{center}       
\begin{tabular}{clcrrrr} 
\hline
 Log Number &\multicolumn{1}{c}{Date} & Exposure & \multicolumn{4}{c}{The equivalent X-ray energy of } \\
     & &  (ks)  & \multicolumn{4}{c}{the injected charge packets (keV)}    \\
     & &          & XIS~0 & XIS~1 & XIS~2 & XIS~3     \\
\hline
  1 &2006/1/17--20 & 3.7 &   -         & 7.1         & -            & 4.6   \\
  2 &2006/5/20--21 & 1.2 & 4.2         &   -         & 3.9          & -   \\
  3 &2006/6/26--27 & 4.9 &   -         &  0.3/7.1    & -            & 0.5/4.5\\  
  4 &2006/7/17     & 5.7 &  0.6/4.2/8.0& -           & 0.6/3.9/7.8  & - \\    
  5 &2006/8/25--26 & 4.9 &   -         & 0.3/7.1     &-             & 0.5/7.0\\
  6 &2007/9/28     &\hspace{-.5em}25.3 & 0.6/4.1/7.9 & 0.4/7.2/7.9 &-             & 0.5/4.5/6.5\\
\hline 
\end{tabular}
\end{center}
\end{table*} 

\begin{table*}
\caption{Observational log of the celestial objects.
\label{tab:objlog}}
\begin{center}       
\begin{tabular}{lclc} 
\hline
 Object & Obs.ID & \multicolumn{1}{c}{Date} & Exposure (ks)      \\
\hline 
The Perseus cluster of galaxies &800010010 & 2006/2/1--2 & 50.4 \\ 
\hline 
1E$\,0102.2-7219$& 100014010 & 2005/8/31      & 24.3 \\
&      100044010 & 2005/12/16--19 & 59.7 \\
&      101005010 & 2006/4/16      & 21.3 \\
&      101005040 & 2006/7/17      & 20.6 \\
&      101005070 & 2006/10/21--22 & 18.5 \\
&      101005100 & 2007/1/15      &  22.6 \\
&      101005120 & 2007/3/18--19  & 18.2 \\                
\hline 
\end{tabular}
\end{center}
\end{table*}

\end{document}